\documentclass[aps,prb,twocolumn,showpacs,superscriptaddress]{revtex4}
\usepackage{bbm}
\usepackage{mathrsfs}
\usepackage{epsfig}
\usepackage{graphicx}
\usepackage{amsfonts}
\usepackage[figuresright]{rotating}
\usepackage{amssymb}
\usepackage{amsmath}
\usepackage{dcolumn}
\usepackage{bm}

\def\avg#1{\langle#1\rangle}

\def\be{\begin{equation}} \def\ee{\end{equation}}
\def\bea{\begin{eqnarray}} \def\eea{\end{eqnarray}}

\def\tr{\mbox{tr}}

\usepackage{ulem}

\usepackage{color}

\begin{document}
\title{Quantum magnetic properties of the SU(2N) Hubbard model
in the square lattice: a quantum Monte Carlo study}
\author{Zi Cai}
\affiliation{Department of Physics, University of California, San
Diego, CA92093} \affiliation{Department of Physics and Arnold
Sommerfeld Center for Theoretical Physics,
Ludwig-Maximilians-Universit{\"a}t M{\"u}nchen, Theresienstr.\ 37,
80333 Munich, Germany}
\author{Hsiang-Hsuan Hung}
\affiliation{Department of Electrical and Computer Engineering,
University of Illinois, Urbana,  Illinois 61801}
\affiliation{Department of Physics, University of California, San
Diego, CA92093}
\author{Lei Wang}
\affiliation{Theoretische Physik, ETH Zurich, 8093 Zurich,
Switzerland}
\author{Congjun Wu}
\affiliation{Department of Physics, University of California, San
Diego, CA92093}

\begin{abstract}
We employ the determinant projector quantum Monte-Carlo method to
investigate the ground state magnetic properties in the Mott
insulating states of the half-filled SU(4) and SU(6) Fermi-Hubbard
model in the 2D square lattice, which is free of the sign problem.
The long-range antiferromagnetic Neel order is found for the SU(4)
case with a small residual Neel moment.
Quantum fluctuations are even stronger in the SU(6) case.
Numeric results are consistent with either a vanishing or even weaker
Neel ordering than that of SU(4).
\end{abstract}
\pacs{71.10.Fd, 02.70.SS, 03.75.Ss, 37.10.Jk,71.27.+a}
\maketitle

Quantum antiferromagnetism (AF) has been an important topic
of the two-dimensional (2D) strongly correlated systems for decades.
For the Hubbard model in the 2D square lattice, charge gap opens
starting from an infinitesimal $U$.
The low energy physics is described by the AF Heisenberg model.
For the SU(2) case, quantum spin fluctuations are not strong enough to suppress
AF long-range order \cite{hirsch1985,hirsch1989}.
Augmenting the symmetry to SU$(N)$ or Sp$(2N)$ enhances quantum spin
fluctuations \cite{arovas1988,affleck1988,sachdev1991},  which can be
handled by the systematic $1/N$-analysis.
The SU($N$) spin operators can be formulated in terms of either bosonic
or fermionic representations.
The bosonic large-$N$ analysis finds gapped quantum paramagnetic
states exhibiting various crystalline orderings \cite{read1989},
while the fermionic one gives rise to gapless flux-type spin liquid states
\cite{affleck1988,hermele2004}.
However, its stability remains an open issue.
On the other hand, short-range resonating-valence-bond type gapped
spin liquid states have also been extensively studied
\cite{anderson1973,rokhsar1988,moessner2001}.

Due to the difficulty of handling strong correlations, numerical
simulations have been playing an important role on the study of
exotic quantum spin states
\cite{harada2003,kaul2012,assaad2005,paramekanti2007,meng2010,
chang2012,yan2011,jiang2011,blumer2013}. Whether the spin-disordered
quantum insulating states exist in the honeycomb lattice or not is
currently under debating \cite{meng2010,sorella2012}. A constrained
path-integral quantum Monte-Carlo (QMC) simulation finds the evidence
of a gapless spin disordered phase in the square lattice with $\pi$-flux per
plaquette \cite{chang2012}.
Evidence of gapped spin liquid phases has also been found by the
density-matrix-renormalization-group simulations of the frustrated
Heisenberg models in the Kagome lattice \cite{yan2011} and in the
square lattice with diagonal couplings \cite{jiang2011}.

The Fermi-Hubbard models with $2N$ components possessing the
SU($2N$) or Sp($2N$) symmetries are not only of academic interest
now, but also have become the goal of experimental efforts in the
ultra-cold atom physics \cite{wu2010}. It was first proposed to use
large-spin alkali and alkaline-earth atoms to realize the Sp($2N$)
and SU($2N$) Hubbard models in Ref. [\onlinecite{wu2003}] for the special
case of $2N=4$ with the proof of a generic Sp$(4)$ symmetry without
fine-tuning. Currently, the SU(6) and SU(10) symmetric systems of
$^{173}$Yb and $^{87}$Sr atoms have been realized, respectively
\cite{gorshkov2010, taie2010,desalvo2010}. In particular, the
$^{173}$Yb atoms have been loaded into optical lattices to realize
the SU(6) Hubbard model, and the charge excitation gap has been
observed \cite{taie2010}. It has also been expected that Pomeranchuk
cooling is efficient in the large-$N$ case to further cool the
system down to the temperature scale of the AF exchanges
\cite{cai2012}.

In this article, we investigate the magnetic properties of the
half-filled SU($2N$) Hubbard models with $2N=4$ and $6$ by the
sign-problem free determinant projector quantum Monte-Carlo (QMC) method.
For the SU(4) case, the ground state remains AF ordered as in the
case of SU(2) although the residual spin moments are much weaker.
For the SU(6) case,  we find that the residual Neel moments are either
absent or extremely small beyond the resolution limit of our
simulations on structure factors and the finite size scaling scheme.

The SU($2N$) Fermi Hubbard model in the 2D square lattice at half-filling
is defined as
\bea
H=-t\sum_{\avg{i,j},\alpha}\Big\{ c^\dag_{i\alpha}c_{j\alpha}+h.c.\Big\}
+\frac{U}{2}\sum_i\big(n_i-N \big)^2,
\label{eq:sun}
\eea
where $t$ is scaled as 1 below; $\alpha$ represents spin indices running
from 1 to $2N$; $\avg{i,j}$ denotes the summation over the nearest
neighbors; $n_i$ is the particle number operator on site $i$ defined as
$n_i=\sum_{\alpha=1}^{2N} c^\dag_{i\alpha}c_{i\alpha}$.
Eq. \ref{eq:sun} is invariant under the particle-hole transformation in
bipartite lattices as $c_{i\alpha}\rightarrow (-)^i c^\dagger_{i\alpha}$,
and thus the average filling per site $\avg{n_i}=N$.
Similarly to the case of SU(2), the SU($2N$) Hubbard model at
half-filling in bipartite lattices is free of the sign problem
for an arbitrary value of $2N$.

We use the determinant projector QMC method for fermions with the
periodical boundary condition \cite{sugiyama1986,white1989,
assaad2008}. The simulated system sizes $L\times L$ range from $L=4$
to $16$. Finite-size scaling is performed to extrapolate the ground
state properties in the thermodynamic limit. The initial trial
wavefunction is the ground state of the free part of Eq.
\ref{eq:sun} whose hopping integral is attached a small flux to
break the degeneracy \cite{meng2010}. Such a Slater-determinant
plane-wave state for the imaginary-time evolution is assumed to be
non-orthogonal to the true ground state of the entire Hamiltonian.
The second order Suzuki-Trotter decomposition is performed with the
imaginary time-step $\Delta \tau=0.05$. The convergence of the
simulation results with respect to different values of $\Delta \tau$
has been checked. The length of the imaginary-time evolution is
$\beta=40$. For the SU(2) case, the Hubbard-Stratonovich (HS)
transformation is usually performed by using the discrete Ising spin
fields \cite{hirsch1985}. However, the spin channel decomposition
does not easily generalize to the $SU(2N)$ case due to the
increasing of spin components. Instead, we follow the approximate
discrete HS decomposition in the density channel at the price of
involving complex numbers  \cite{assaad1998}. The error of this
approximation is at the order $(\Delta \tau)^4$, smaller than that
of the Suzuki-Trotter decomposition, thus is negligible. This method
has the advantage that the SU($2N$) symmetry is maintained
explicitly, and also it easily generalizes to large values of $2N$.

Let us fix the convention of the SU($2N$) generators.
The Hilbert space on site $i$ filled with $r~(1\le r \le 2N)$ fermions
forms the $SU(2N)$ representation described by the single column Young pattern
denoted as $1^r$ where $r$ is the number of rows.
For these $1^r$-representations, the $SU(2N)$ generators are defined as
\bea
J^{\alpha\beta}(i)=c^\dag_{\alpha}(i)c_{\beta}(i)- \frac{\delta^{\alpha\beta}}
{2N}\sum_{\gamma=1}^{2N} c^\dag_\gamma(i) c_\gamma(i).
\label{eq:su2n_gen}
\eea
Another standard definition is through the generalized Gell-mann matrices
$c^\dagger_\alpha(i) \lambda^a_{\alpha\beta} c_\beta(i)$ with $ 1\le a \le 4N^2-1$
and the normalization condition of $\tr [\lambda^a \lambda^b]=
\frac{1}{2}\delta^{ab}$.
The definition in Eq. \ref{eq:su2n_gen} has a simple commutation
relation as
$[J^{\alpha\beta}, J^{\gamma\delta}]=\delta_{\beta\gamma} J^{\alpha\delta}
-\delta_{\alpha\delta} J^{\gamma\beta}$.
However, the price is that not all of the operators of
Eq. \ref{eq:su2n_gen} are independent, which
satisfy the constraint $\sum_{\alpha} J^{\alpha\alpha}=0$.
The quadratic Casimir operator is
expressed as
$C_{2}(2N)=\frac{1}{2}\sum_{\alpha\beta} J^{\alpha\beta}(i) J^{\beta\alpha}(i)$.
For the $1^r$ representation denoted by the Young pattern with
a single coloumn with $r$ boxes,
its value is related to the filling number $r$ through the Fierz identity as
$
C_2(2N,r)= r(2N-r)(2N+1)/(4N).
$
In the large $U$ limit in which charge fluctuations are negligible,
each site represents the self-conjugate representation $1^N$.
The two-site equal time spin-spin correlation function is defined as
\bea
C_{J,SU(2N)}(i,j)= \frac{1}{C_2(2N,N)} \sum_{\alpha,\beta}
\frac{1}{2} \langle J^{\alpha\beta}(i) J^{\beta\alpha}(j) \rangle,
\label{eq:SS}
\eea
where $C(2N,N)=N(2N+1)/4$ is the Casimir for $1^N$ representation.
$C_{J,SU(2N)}(i,i)$ approaches 1 in the large $U$-limit.
The normalized spin structure factor at the AF wavevector $\vec Q$ is
defined as
\bea
S_{SU(2N)}(\vec Q)
&=& \frac{1}{C_2(2N,N)} \sum_{\alpha\beta}
 \frac{1}{2}\avg{J^{\alpha\beta}(\vec Q) J^{\beta\alpha}(\vec Q)}, \ \ \
\eea
where $J^{\alpha\beta}(\vec Q)=\frac{1}{L} \sum_i e^{i \vec Q \cdot \vec r_i}
J^{\alpha\beta}(i)$.
The imaginary-time-displaced spin-spin correlations at
wavevector $\vec Q$ are defined as
\bea
S_{SU(2N)}(\vec Q,\tau)=\sum_{\alpha\beta}\avg{J^{\alpha\beta}(\vec Q, \tau)
J^{\beta\alpha}(\vec Q,0)},
\eea
which are used to extract spin gaps below.

\begin{figure}[htb]
\includegraphics[width=0.85\linewidth]{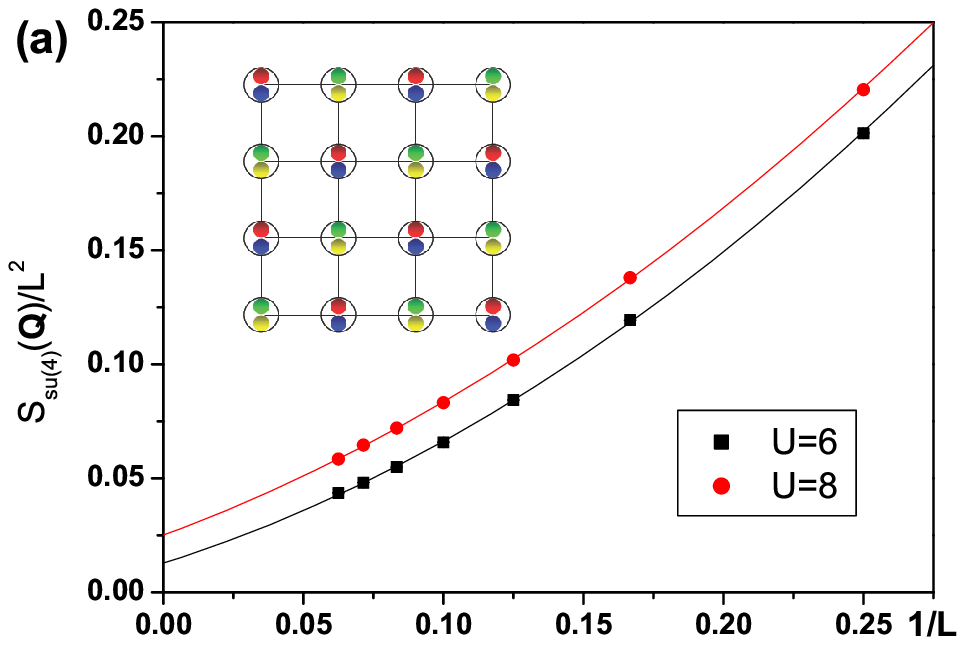}
\includegraphics[width=0.85\linewidth]{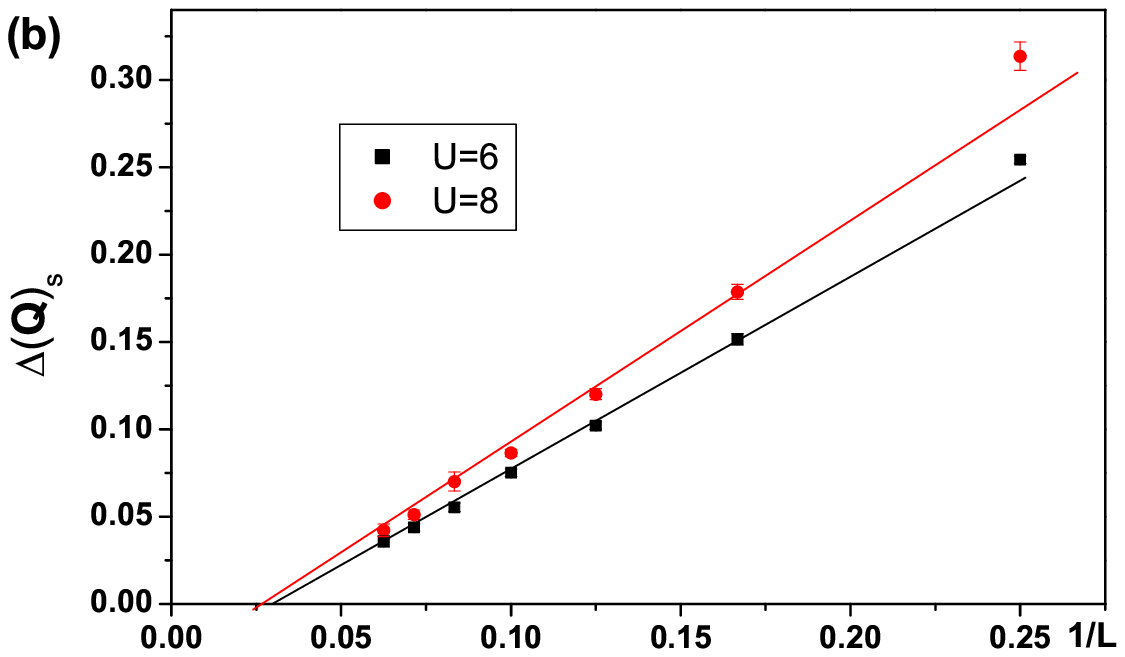}
\caption{The half-filled SU(4) Hubbard model in the square lattice.
(a) The appearance of the AF long-range-order from the finite size
scaling of the spin structure factor at $\vec Q=(\pi,\pi)$ for $U=6$
and $8$. Solid curves are quadratic fits of data. The inset shows a
typical SU(4) AF configuration in which different colors represent
different spin components. (b) The absence of the spin gap from the
finite size scaling of $\Delta_{s}(\vec Q)$. } \label{fig:su4spin}
\end{figure}

{\it The SU(4) case~~} Below we present the study of quantum spin
fluctuations starting with the $SU(4)$ case in the square lattice,
in which we find long-range AF ordering since intermediate values of
$U$. The finite size scaling of the spin structure factor
$\frac{1}{L^2} S_{SU(4)}(\vec Q)$ at the AF wavevector $\vec
Q=(\pi,\pi)$ is plotted in Fig. \ref{fig:su4spin} (a). For example,
at $U=8$, it extrapolates to a small but finite value of $s_0=0.025$
as $L\rightarrow \infty$, which indicates the existence of the AF
long-range Neel order. In comparison, for the SU(2) case at the same
value of $U$, the extrapolated value of $\lim_{L\rightarrow \infty}
\frac{1}{L^2} S_{SU(2)}(\vec Q) \approx 0.118$. This shows the
enhancement of quantum spin fluctuations as $2N$ increases.

Let us bipartite the lattice into $A$ and $B$ sublattices.
One typical classic SU(4) Neel configuration is that $A$-sites are filled
with components 1 and 2, and $B$-sites filled with components 3 and 4.
SU(4) is a rank-3 Lie group, and thus its Cartan algebra has three
commutable generators defined as
$K_{1,2}=\frac{1}{2\sqrt 2} [(n_1-n_2) \pm (n_3-n_4)]$,
and $K_3=\frac{1}{2\sqrt 2} [(n_1+n_2) -(n_3+n_4)]$.
Each site of the above $SU(4)$ configuration is a singlet of
$K_{1,2}$, and with the eigenvalues of $\pm \frac{1}{\sqrt 2}$ for $K_3$.
The AF long-range-ordered states possess gapless Goldstone modes,
and the Goldstone manifold is the 8-dimensional Grassmann
one $U(4)/[U(2)\times U(2)]$.
The spin excitations carry quantum numbers of $K_{1,2,3}$
as $(\pm \frac{1}{\sqrt 2}, 0, \pm \frac{1}{\sqrt 2})$ and
$(0, \pm \frac{1}{\sqrt 2}, \pm \frac{1}{\sqrt 2})$.
To verify the absence of spin gap, we calculate the imaginary-time-displaced
spin correlation function $S_{SU(4)}(\vec Q, \tau)$
\cite{assaad1996,feldbacher2001}.
The finite size spin-gap $\Delta_s(\vec Q, 1/L)$ is fitted from the slope of
$\ln S_{SU(4)} (\vec Q,\tau)$ v.s. $\tau$.
The finite-size scaling is plotted in Fig. \ref{fig:su4spin} (b) which shows
the absence of spin gap in consistent with the long-range AF ordering.

\begin{figure}[htb]
\includegraphics[width=0.85\linewidth]{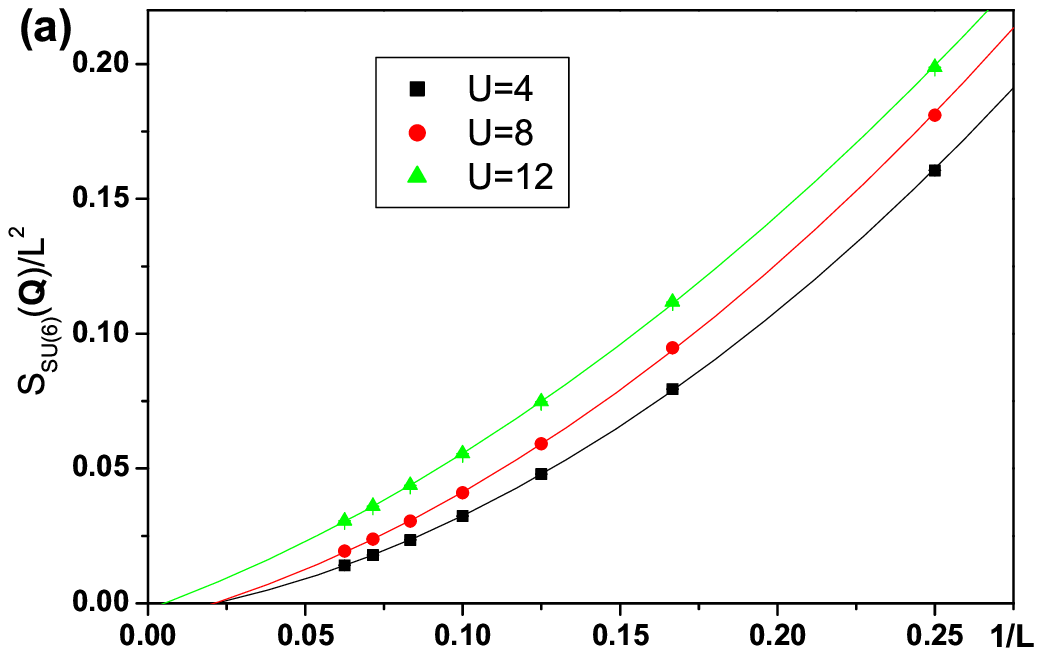}
\includegraphics[width=0.85\linewidth]{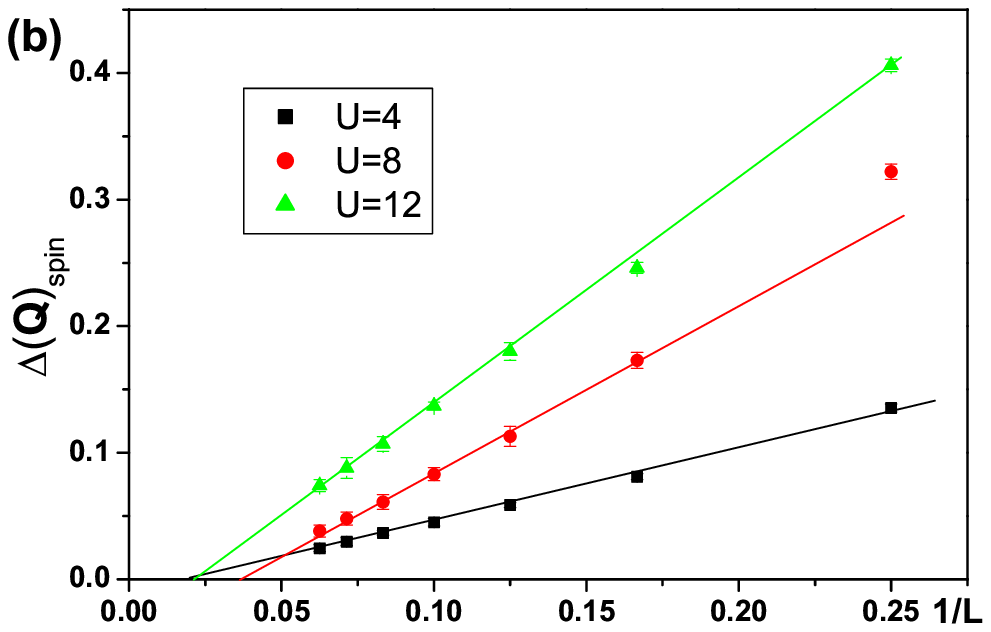}
\includegraphics[width=0.85\linewidth]{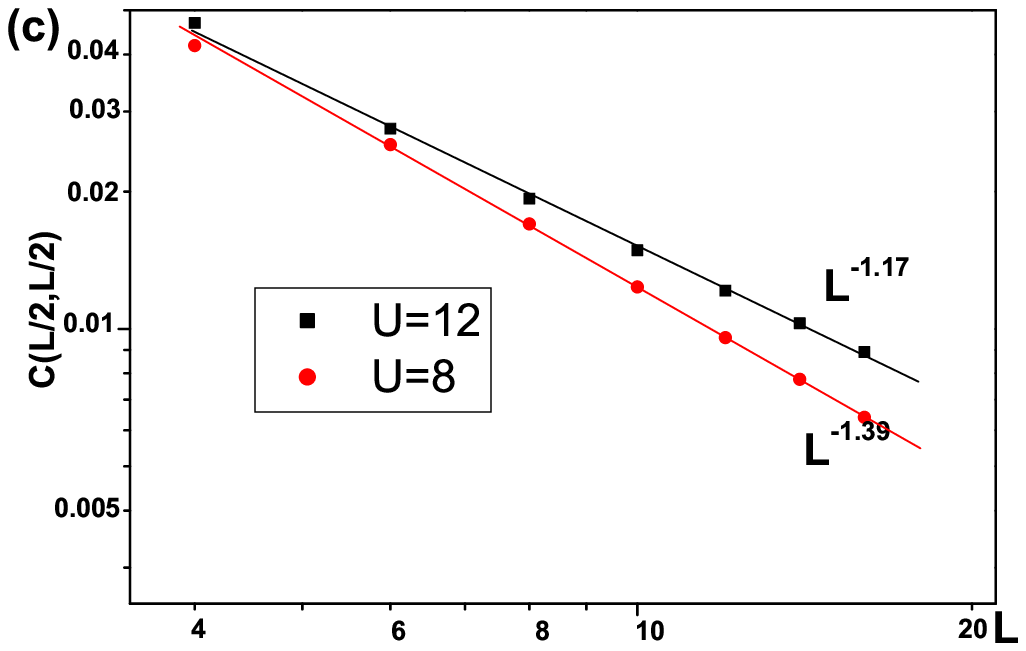}
\caption{Spin correlations of the half-filled SU(6) Hubbard model.
(a) The finite-size scalings of the spin structure factors at $\vec
Q=(\pi,\pi)$ at $U=4,8$ and 12 are consistent with either zero
or a very weak Neel ordering. Solid curves are quadratic fits.
(b) The finite-size scalings
$\Delta_{s}(\mathbf{Q})$ show the absence of spin gap. (c) The
scalings of the farthest point correlations $C_{J, SU(6)}(L/2,L/2)$
for $U=8$ and 12. } \label{fig:su6spin}
\end{figure}

{\it The SU(6) case~~} As $2N$ increases to 6, quantum spin fluctuations
become even stronger. The QMC simulation of the  spin structure factors
at $\vec Q=(\pi,\pi)$ is presented in Fig. \ref{fig:su6spin} (a).
The finite size scalings of the SU(6) AF structure factor for all the
cases of $U=4,8$ and $12$ extrapolate to zero.
However, because the $1/L$ extrapolation of the AF structure factor
is proportional to the square of the AF moments, the possibility of a
weak AF long-range-order cannot be excluded.
For example, a Neel moment at the order of $10^{-2}$ corresponds
to the structure factor at the order of $10^{-3}$ or $10^{-4}$,
which is beyond our current resolution limit.
We further calculate the spin gap value at $\vec Q=(\pi, \pi)$ from
the imaginary-time-displaced $SU(6)$ spin correlation function
$S_{SU(6)}(\vec Q, \tau)$, and plot the extracted spin gap values in
Fig. \ref{fig:su6spin} (b).
The finite-size scaling shows the vanishing of spin gap in the SU(6) case
for all the three values of $U=4,8$ and $12$.
The vanishing of spin gaps are also consistent with very small but nonzero
AF moments.
The two-point equal-time spin-spin correlations $C_{J,SU(6)}(L/2,L/2)$
are calculated and plotted in Fig. \ref{fig:su6spin} (c), which are
fitted with algebraic correlations as $C_{J,SU(6)} (L/2,L/2)
\approx L^{-\eta}$.
However, due to the limited sample size, these algebraic correlations
are well fitted at a intermediate length scale.
We still cannot exclude the possibility of small long-range AF moments.

\begin{figure}[htb]
\includegraphics[width=0.7\linewidth]{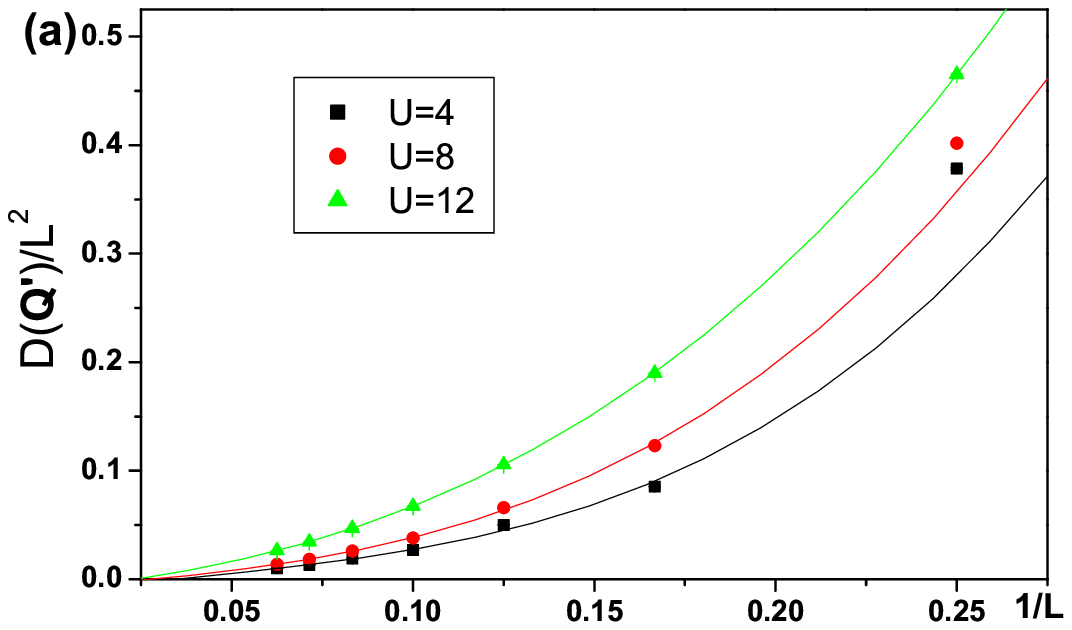}
\includegraphics[width=0.7\linewidth]{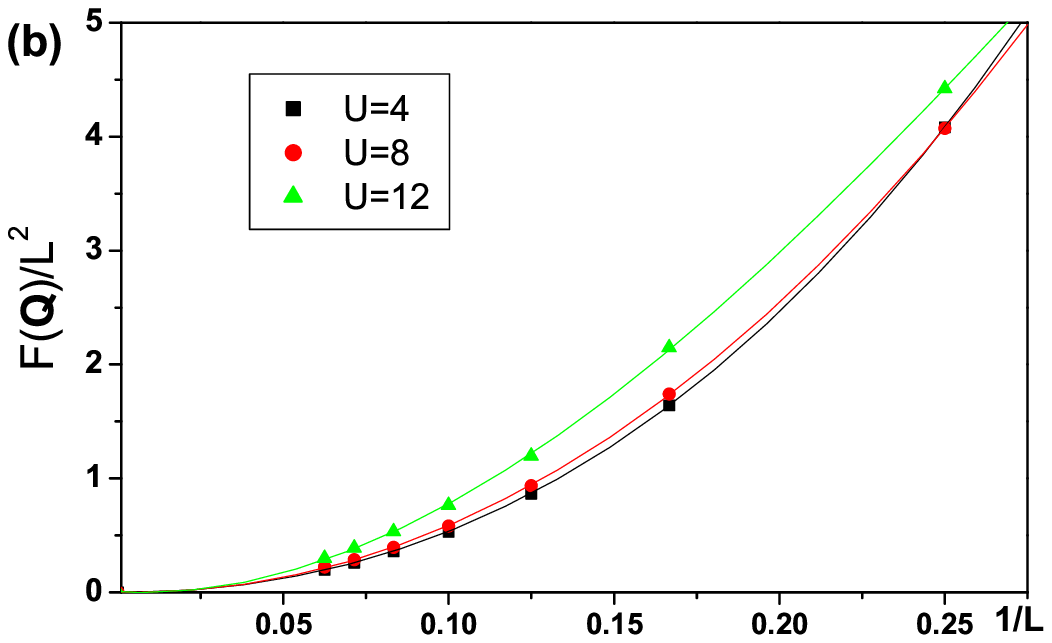}
\caption{Spin singlet channel operators of the half-filled SU(6) Hubbard model.
(a) The finite-size scaling of the columnar dimer structure factors at
$\vec Q^\prime=(\pi,0)$.
(b) The finite-size scaling of the DDW structure factors at
$\vec Q=(\pi,\pi)$.
}
\label{fig:su6-2}
\end{figure}

We further check other possible ordering patterns involving two
neighboring sites.
At half-filling, total particle number on a bond is $2N$, which is sufficient
to form an SU(2N) singlet to minimize the spin superexchange energy.
We consider ordering patterns in the spin singlet channel with translational
symmetry breaking.
The bond dimer and current operators are defined as
the real and imaginary parts of the hopping amplitudes between
nearest neighbors as
\bea
D_{ij}= \sum_\alpha c^\dag_{i,\alpha}c_{j,\alpha} +h.c., \ \
F_{ij}= \sum_\alpha i(c^\dag_{i,\alpha}c_{j,\alpha} -h.c.), \ \ \
\eea
and $d$-density-wave (DDW) operators as $DDW(i)=(-)^i \sum_j F(i, j)$
where $\vec r_j -\vec r_i= \pm \hat e_x$, and $\pm \hat e_y$.
In the large $U$ limit, the Heisenberg term $S^{\alpha\beta}(i)S^{\beta\alpha}(j)$
is generated from the second order virtual hopping process, thus
$D_{ij}$ can be used as the dimer order parameter.
The structure factor of $D_{ij}$ at $\vec Q^\prime=(\pi,0)$ and
that of DDW at $\vec Q=(\pi,\pi)$, after divided by $L^2$,
and are plotted in Fig. \ref{fig:su6spin} b) and c), respectively.
They are fitted by a power-law $(1/L)^2$, thus their correlations
are short-ranged.

\begin{figure}[htb]
\includegraphics[width=0.485\linewidth]{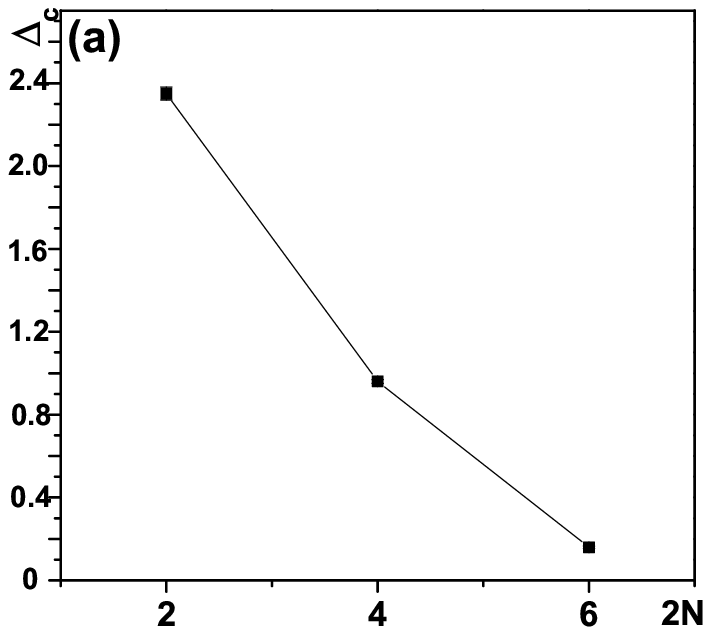}
\includegraphics[width=0.495\linewidth]{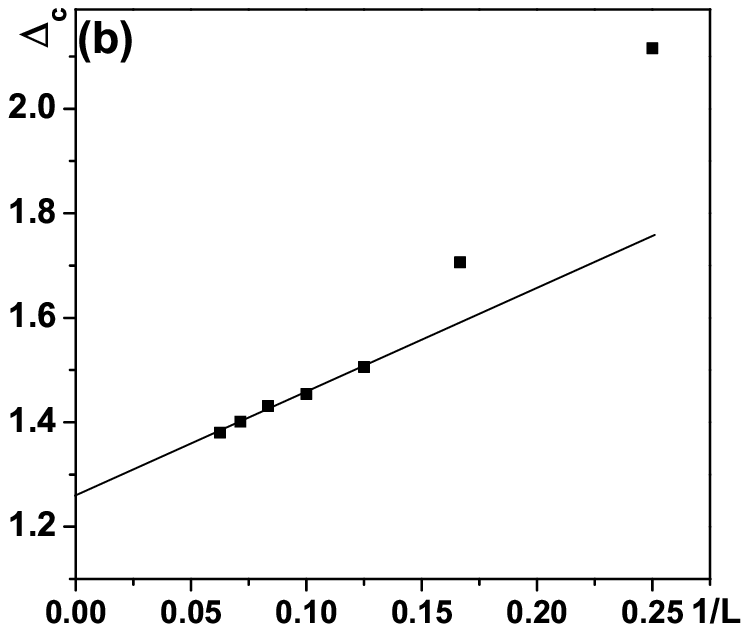}
\caption{Single-particle gaps of half-filled SU($2N$) Hubbard models. A)
Charge gaps with $U=8$ at $2N=2,4$ and $6$. B) The $1/L$ scaling of
the charge gap for the half-filled SU(6) model at $U=12$. }
\label{fig:charge_gap}
\end{figure}

{\it Single-particle gaps} The single-particle gaps for the SU(4)
and SU(6) Hubbard models are also calculated at half-filling through
the onsite imaginary time-displaced Green's function
$G(0,\tau)=\frac{1}{L^2} \sum_i \avg{\Psi_G|c(i,\tau)
c^\dagger(i,0)|\Psi_G}$, where $|\Psi_G\rangle$ is the ground state.
At long time displacement, $G(0,\tau)\rightarrow e^{-\Delta_c\tau}$
where $\Delta_c$ is the single-particle excitation gap, thus
$\Delta_c$ can be fitted from the slope of $\ln G(0,\tau)$ v.s.
$\tau$. Let us consider the large $U$-limit for an intuitive
picture: in the Mott-insulating background, the energy of adding a
particle is lowered from $U$ by further virtual particle-hole
excitations. In other words, the Mott-insulator is polarizable. As
increasing $2N$, the configuration numbers of the virtual
particle-hole excitations increase, which enhances charge
fluctuations and thus reduces the single-particle gap. In Fig.
\ref{fig:charge_gap} (a), $\Delta_c$'s are plotted at a fixed $U=8$
for $2N=2,4$ and $6$, all of which are finite.
For the SU(6) case, $\Delta_c=0.15$ is rather small at $U=8$.
Nevertheless, $\Delta_c$ increases to $1.26$
at $U=12$ at which the system is safely inside the Mott-insulating
regime. The charge localization length can be estimated as
$\xi_c\approx v_f/\Delta_c \approx 3\sim 4$ which is much smaller
than the maximal sample size $L=16$.

In conclusion, we have studied the ground state quantum antiferromagnetism
in a half-filled SU($2N$) Hubbard model in square lattice.
For the case of SU(4), a long-range AF order still survives with a much
smaller value of Neel moment compared to that of SU(2).
For the SU(6) case, we have found the absence of spin gap.
The current numeric results are consistent with either a vanishing
or very weak AF ordering beyond the resolution limit in this simulation.
We have also found that the single particle gap is strongly
suppressed as increasing $N$.

C. W. thanks J. Hirsch and S. Kivelson for very helpful discussions.
Z.C thanks F.F. Assaad for helpful discussion. Z. C., Y. L. and C.
W. are supported by the NSF DMR-1105945 and the AFOSR
FA9550-11-1-0067(YIP program).




\end{document}